\documentclass[preprint,pdf]{iucr} 
\RequirePackage{graphicx}
\journalcode{B}
\usepackage{graphicx}
\usepackage{epsfig}
\usepackage{amsmath,amssymb}
\usepackage{bm}
\usepackage{ cleveref}

\begin{document}
	\title{Direct interpretation of the X-ray and neutron 3D-$\Delta$PDFs of yittria stabilized zirconia}
	\cauthor[a,b]{Ella Mara}{Schmidt}{ella.schmidt@uni-bremen.de}{}
    \author[c]{Reinhard B.}{Neder}
    \author[d]{James D.}{Martin}
    \author[b]{Arianna}{Minelli}
    \author[e]{Marie-Hélène}{Lemée}
    \author[b]{Andrew L.}{Goodwin}
    \aff[a]{Faculty of Geosciences, MARUM and MAPEX, University of Bremen, Bremen, Germany}
    \aff[b]{Inorganic Chemistry Laboratory, University of Oxford, Oxford, United Kingdom}
    \aff[c]{Institute of Condensed Matter Physics, Friedrich-Alexander University, Erlangen, Germany}
    \aff[d]{Department of Chemistry, North Carolina State University, Rayleigh, USA}
    \aff[e]{Institute Laue Langevin, Grenoble, France}

	\maketitle                        
	
	\begin{synopsis}
We use 3D-$\Delta$PDFs from X-ray and neutron diffraction experiments to identify local stabilization mechanisms in yttria stabilized cubic zirconia.
	\end{synopsis}
	
	\begin{abstract}
		
		Three dimensional difference pair distribution functions from X-ray and neutron diffraction experiments are reported for yttria stabilized zirconia (Zr$_{0.82}$Y$_{0.18}$O$_{1.91}$).  We use a quantitative analysis of the signatures in the 3D-$\Delta$PDFs to establish that oxygen ions neighbouring a vacancy shift by 0.515(5)~\AA\ along $\langle 1,0,0 \rangle$ towards the vacancy while metal ions neighbouring a vacancy shift by 0.269(2)~\AA\ along $\langle 1,1,1 \rangle$ away from the vacancy. The neutron 3D-$\Delta$PDF shows a tendency for vacancies to cluster along $\langle \frac{1}{2},\frac{1}{2},\frac{1}{2}\rangle$, which results in 6-fold coordinated metal ions.

	\end{abstract}

	\section{Introduction}
	
	The technological importance of zirconia (ZrO$_2$) is undoubted and underlined by its numerous applications in the ceramic industry.
	At ambient pressure and temperature ZrO$_2$ is monoclinic ($P2_{1}/c$) but upon increasing temperature it transforms to a tetragonal phase ($P4_{2}/nmc$, at approximately 1440~K) and a cubic phase ($Fm\bar{3}m$, at approximately 2640~K) \cite{boysen1991neutron,bondars1995powder}.
	The addition of aliovalent oxides (CaO, MgO or Y$_{2}$O$_{3}$) allows the stabilization of the cubic phase at ambient conditions  and yields a material that is widely applied because of its strength, high refractive index and thermal-shock resistance \cite{stapper1999ab}.
	To maintain the overall charge balance in cubic stabilized zirconia (CSZ), oxygen vacancies are introduced into the structure which makes CSZ an ion conductor used in solid oxygen fuel cells \cite{tsampas2015applications}, oxygen sensors \cite{schindler1989spectroscopic} and oxygen pumps \cite{pham1998oxygen}.

\begin{figure}
	\caption{Schematics of local relaxations in YSZ as reported in literature \cite{frey2005diffuse, asakib1998cation, fevre2005local}. Oxygen ions (red) relax along  $\langle 1, 0, 0 \rangle$ towards a neighbouring vacancy, next neighbour (NN) Zr ions relax along $\langle 1, 1, 1 \rangle$ away form the vacancy, while  next nearest neighbour (NNN) dopant ions do not show significant relaxations.}
	\includegraphics[width=9cm]{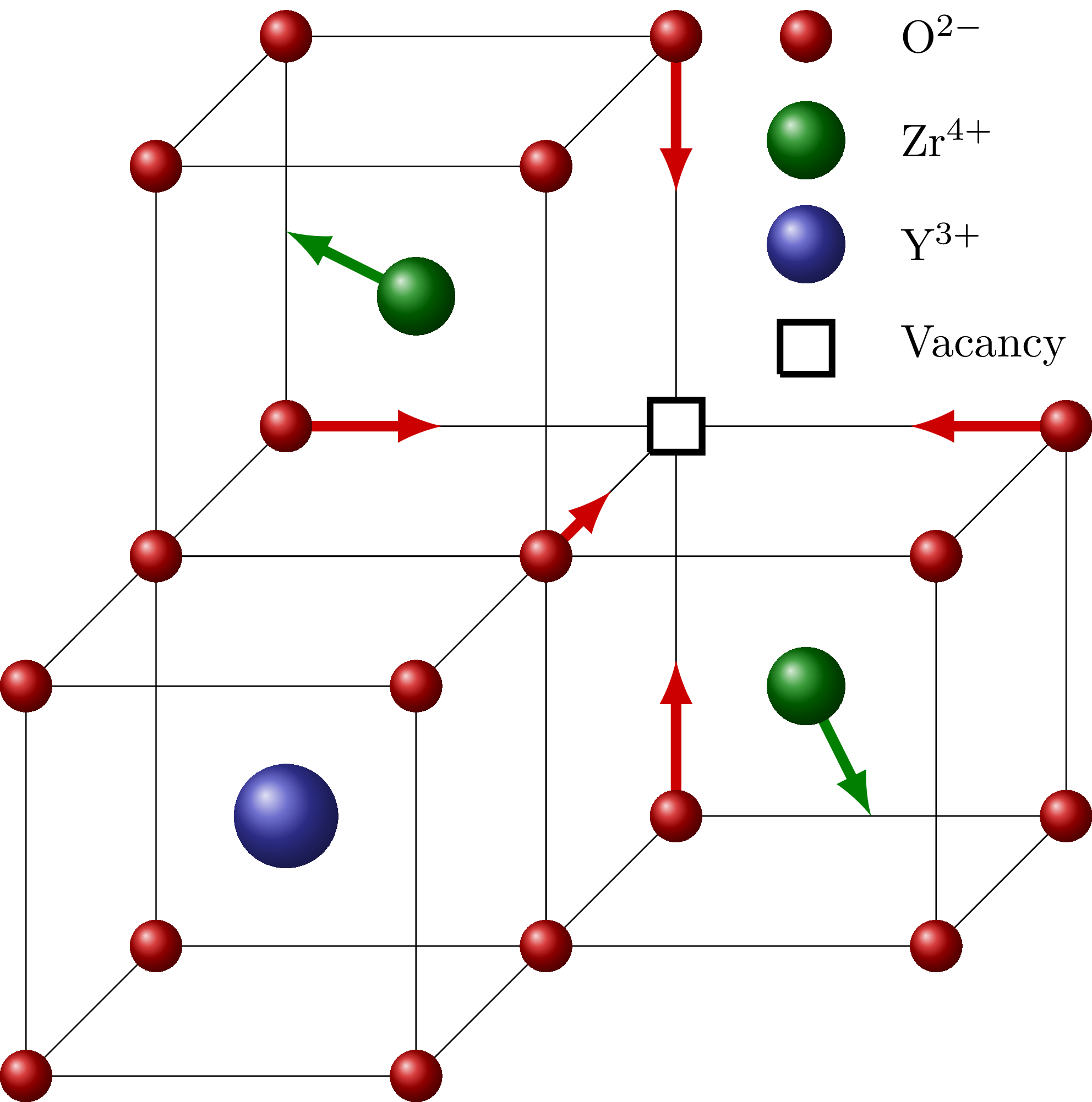}
	\label{Fig:Relaxation}
\end{figure}

   The high temperature cubic polymorph of pure ZrO$_{2}$ adapts the fluorite structure, where each Zr$^{4+}$ is in regular 8-fold coordination. 
   Upon cooling, the local coordination environment in pure ZrO$_{2}$ is distorted and Zr$^{4+}$ is seven-fold coordinated.
   On average, CSZ still adapts the fluorite structure but the introduction of oxygen vacancies ($\square$) reduces the average coordination number to $8-x$, where $x$ depends on the dopant ion charge and concentration.
	The resulting complex defect structure of CSZ is a long standing problem that has been addressed by numerous computational and experimental means. 
	The nature of the most localized interactions is widely agreed upon in literature and reproduced in Figure~\ref{Fig:Relaxation} \cite{frey2005diffuse, asakib1998cation, fevre2005local}: oxygen ions neighbouring a vacancy relax along $\langle 1, 0, 0 \rangle$ towards this  vacancy.
	Electrostatic effects and atomic sizes need to be taken into account when evaluating which type of metal ions neighbours a vacancy.
	Whether the oxygen vacancies are located preferably as a nearest neighbour (NN, i.e. in the coordination of the dopant metal ion) or a next nearest neighbour (NNN) depends on the concentration, size and charge of the dopant ion \cite{asakib1998cation}.
	For YSZ, simulations show that NNN vacancies are preferred \cite{bogicevic2003nature, asakib1998cation}.
	The Zr$^{4+}$ ions next to the vacancy relax along $\langle 1, 1, 1 \rangle$ away from the  vacancy \cite{frey2005diffuse, asakib1998cation, fevre2005local}. The exact magnitude of the relaxations and correlated further neighbour displacement depend on type and concentration of the dopant ion and only a limited agreement is reported in literature (see \citeasnoun{frey2005diffuse} for a review).
	
	In our contribution we focus on the defect structure in yttria stabilized zirconia (YSZ) because of its technological importance as an oxygen ion conductor. We use 3D-$\Delta$PDFs from neutron and X-ray diffraction experiments to quantify local order principles.
	The material under investigation is by no means a newly studied material - YSZ has been first described in 1951 by \citeasnoun{hund1951anomale}. Previous theoretical and experimental investigations include first principle calculations \cite{stapper1999ab}, molecular dynamics simulations \cite{bogicevic2003nature,fabris2002stabilization},  Bragg data analysis \cite{kaiser2005anion,ishizawa1999synchrotron,morinaga1979x}, extended X-ray absorption fine structure (EXAFS)  \cite{ishizawa1999synchrotron,catlow1986exafs,veal1988exafs}, NMR measurements \cite{viefhaus2006solid,kim2007high} and single crystal diffuse scattering analysis \cite{andersen1986defect,welberry1992diffuse,welberry1995modulation,goff1999defect}. 
	 The latter has so far been restricted to selected layers in reciprocal space but more recent computational and technological advances allow the collection and interpretation of full three dimensional data \cite{welberry2003high} and hence open the possibility to re-visit controversially discussed defect clusters by utilizing the highly powerful 3D-$\Delta$PDF technique \cite{weber2012three,roth2019solving,simonov2014experimental}.
	
	This work is organized as follows. We begin by presenting our experimentally obtained diffuse scattering and quantitatively analyse the 3D-$\Delta$PDFs for the dominant local interactions. We compare these results to literature findings and discuss the novel insights by the 3D-$\Delta$PDF analysis.
	Using our 3D-$\Delta$PDF analysis we perform Monte Carlo simulations to show how the quantitative insight into local order models can be realized in a model crystal.
	We conclude by discussing the significance of our quantitative 3D-$\Delta$PDF analysis and outline the strengths of combining X-ray and neutron single crystal diffuse measurements to solve complex local order problems as encountered in YSZ.

\section{Results and Discussion}

\subsection{Sample material}
The zirconia samples have a composition of  Zr$_{0.82}$Y$_{0.18}$O$_{1.91}$, grown by the skull melting method, delivered by Djevahirdjan S. A., Monthey, Switzerland.
The composition was confirmed by EDX measurements (see supporting information).
For neutron measurements the large, clear single crystals were cut with a diamond saw to cubes with an edge length of approximately 5~mm.
For X-ray diffraction measurements the larger crystals were mechanically ground to a diameter of about 150~$\mu$m and polished.
All measurements were carried out at ambient conditions.

    \begin{figure}
    \caption{Reconstructed and symmetry averaged diffuse scattering of YSZ obtained from neutron (left) and X-ray (right) diffraction experiments. (a) $hk0$-layer, (b) $hhl$-layer.}
    \includegraphics[width=9cm]{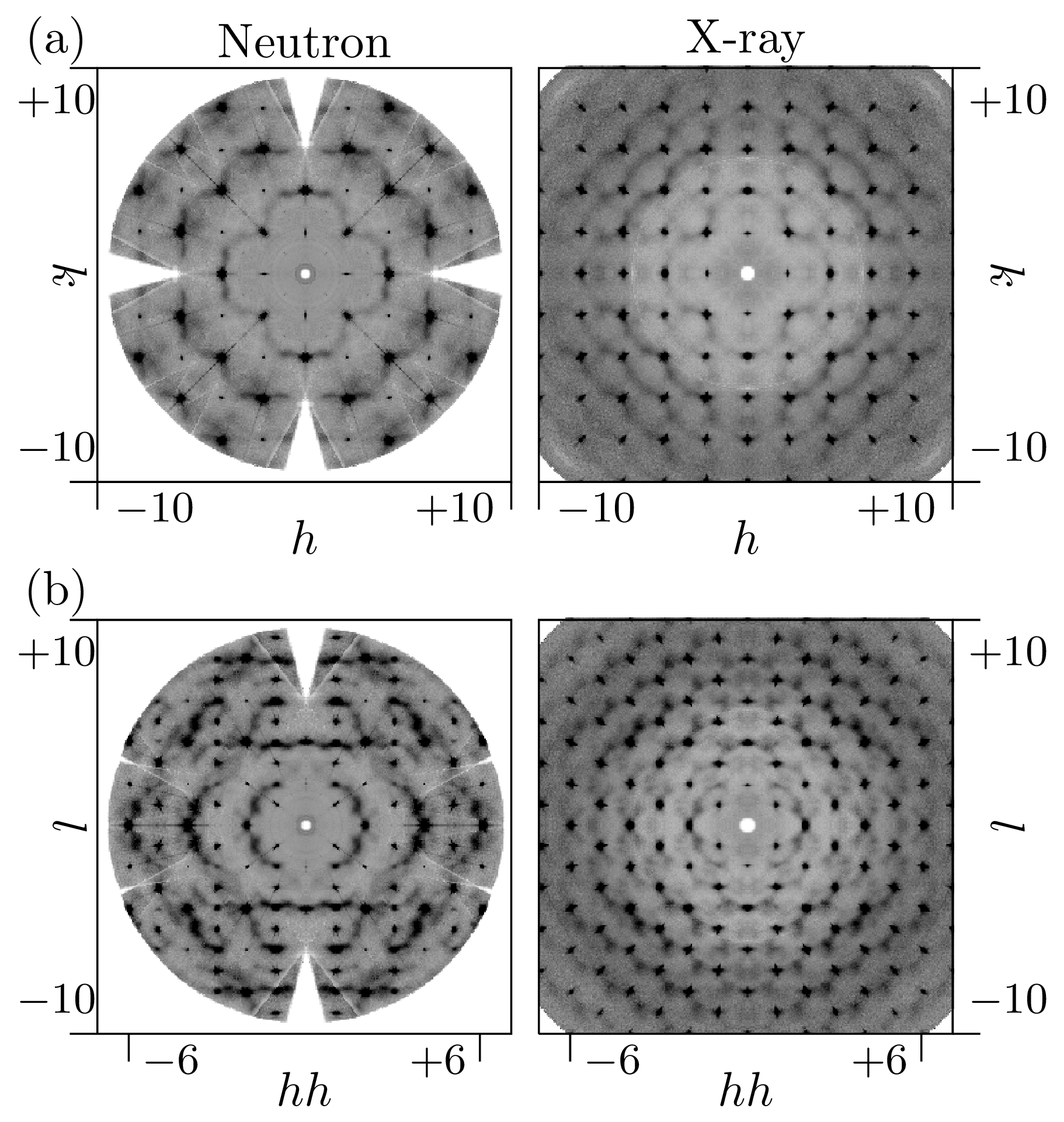}
    \label{Fig:ReciprocalMaps}
\end{figure}

\subsection{X-ray and neutron diffraction experiments}
X-ray diffraction experiments were performed on a Rigaku Synergy S diffractometer equipped with an Eiger 1M detector using Mo radiation.
To avoid possible fluorescence a threshold of 17.4 keV was used on the detector.
Simple $\phi$-scans with 0.5$^{\circ}$ step widths and 120~s exposure time were taken.
3D diffuse scattering data was reconstructed using the orientation matrix provided by CrysAlis Pro \cite{crysalis} and custom Python scripts using Meerkat \cite{meerkat}.

Neutron diffraction experiments were carried out at D19 ($\lambda$ = 0.95~\AA, 0.1$^{\circ}$ steps, 80~s exposure per frame), ILL, Grenoble utilizing a 180$^{\circ}$ $\phi$ scan. 3D diffuse scattering data reconstruction utilized the orientation matrix as provided by Int3d \cite{katcho2021int3d} and a custom Python script.

The reflection conditions for the $Fm\bar{3}m$ space group were fulfilled in all cases and after careful inspection the data were symmetry averaged for $m\bar{3}m$ Laue symmetry. Symmetry averaged reciprocal space maps of selected layers are presented in \Cref{Fig:ReciprocalMaps}.

\subsection{3D-$\Delta$PDF maps}

The general data processing procedure to obtain 3D-$\Delta$PDF experiments is described in \citeasnoun{koch2021single}.
The experimentally obtained data were treated with the KAREN outlier rejection algorithm \cite{weng2020k} and additionally a custom punch-and-fill approach that interpolates the intensity in punched voxels was used to eliminate residual Bragg intensities.
To avoid Fourier ripples the data were multiplied with a Gaussian falloff that smooths the edges of the measured reciprocal space section (see \citeasnoun{weng2020k}).
The FFT algorithm as implemented in Meerkat \cite{meerkat} was used to obtain 3D-$\Delta$PDF maps.

\begin{figure}
    \caption{3D-$\Delta$PDF maps of YSZ obtained from neutron (left) and X-ray (right) diffraction experiments. (a) $ab0$-layer, (b) $aac$-layer. Positive intensities in red, negative intensities in blue.}
    \includegraphics[width=9cm]{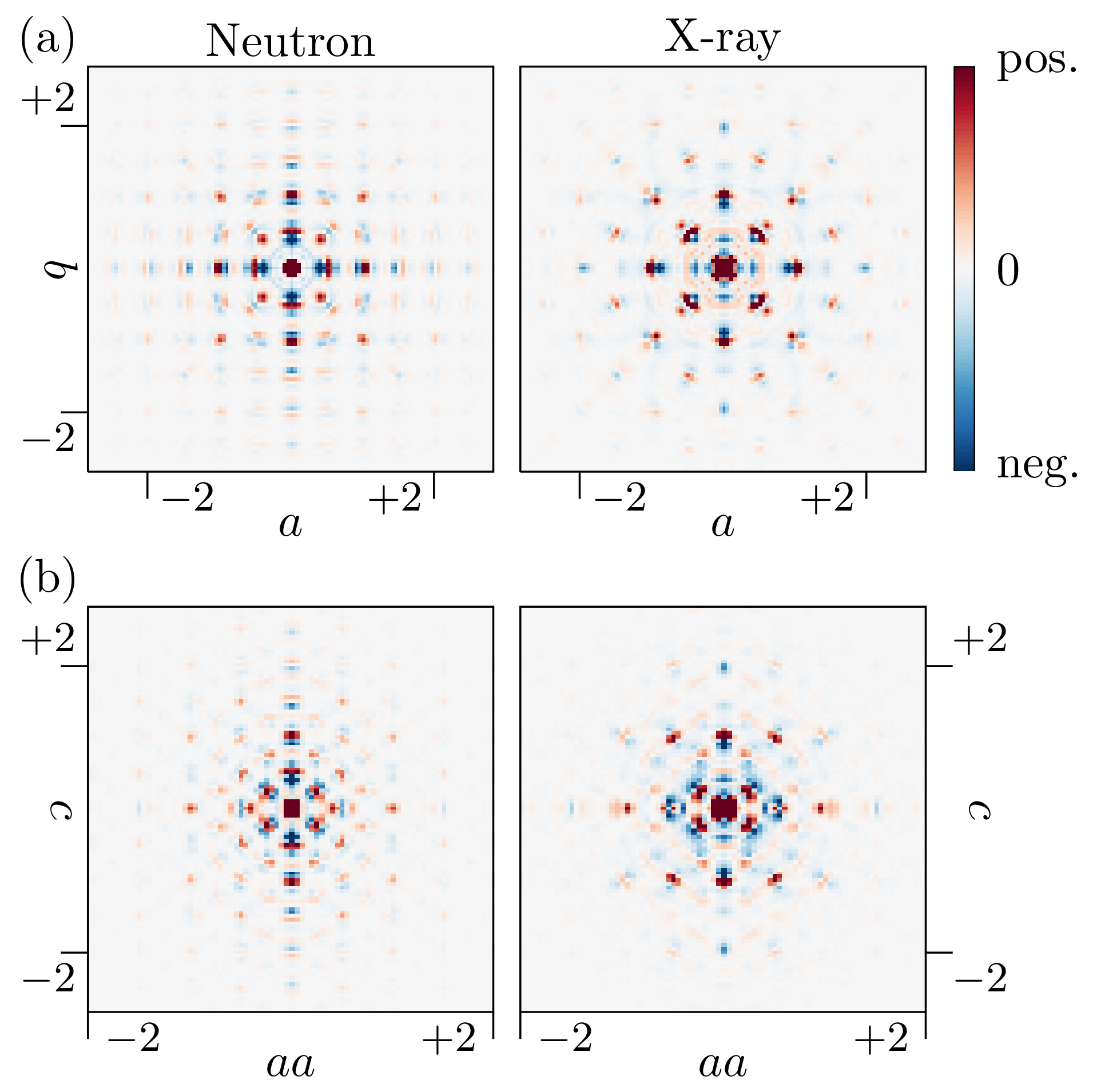}
    \label{Fig:RealSpaceMaps}
\end{figure}

The 3D-$\Delta$PDF maps in the $ab$0-layer and the $aac$-layer are shown in \Cref{Fig:RealSpaceMaps}.
All 3D-$\Delta$PDFs show a variety of different signatures.
The most pronounced features are observed at the shortest interatomic distances indicating strong local order principles, which we will analyse in the following sections.

\subsection{ Oxygen nearest neighbour interactions}

\begin{figure}
    \caption{3D-$\Delta$PDFs of YSZ obtained from neutron (left) and X-ray (right) diffraction experiments around the $\left( \frac{1}{2}, 0, 0 \right)$ interatomic vector. (a) Two-dimensional section in the $ab0$-layer. (b) Three-dimensional rendering of the intensity distribution. Volume shown in the region  $-0.15 \le a - \frac{1}{2},b,c \le 0.15$. Black lines indicate the average structure interatomic distance at $\left( \frac{1}{2}, 0, 0 \right)$. 3D renderings are on the same relative intensity scale with respect to the minimum intensity in the rendering section. Positive intensities in red, negative intensities in blue.}
    \includegraphics[width=9cm]{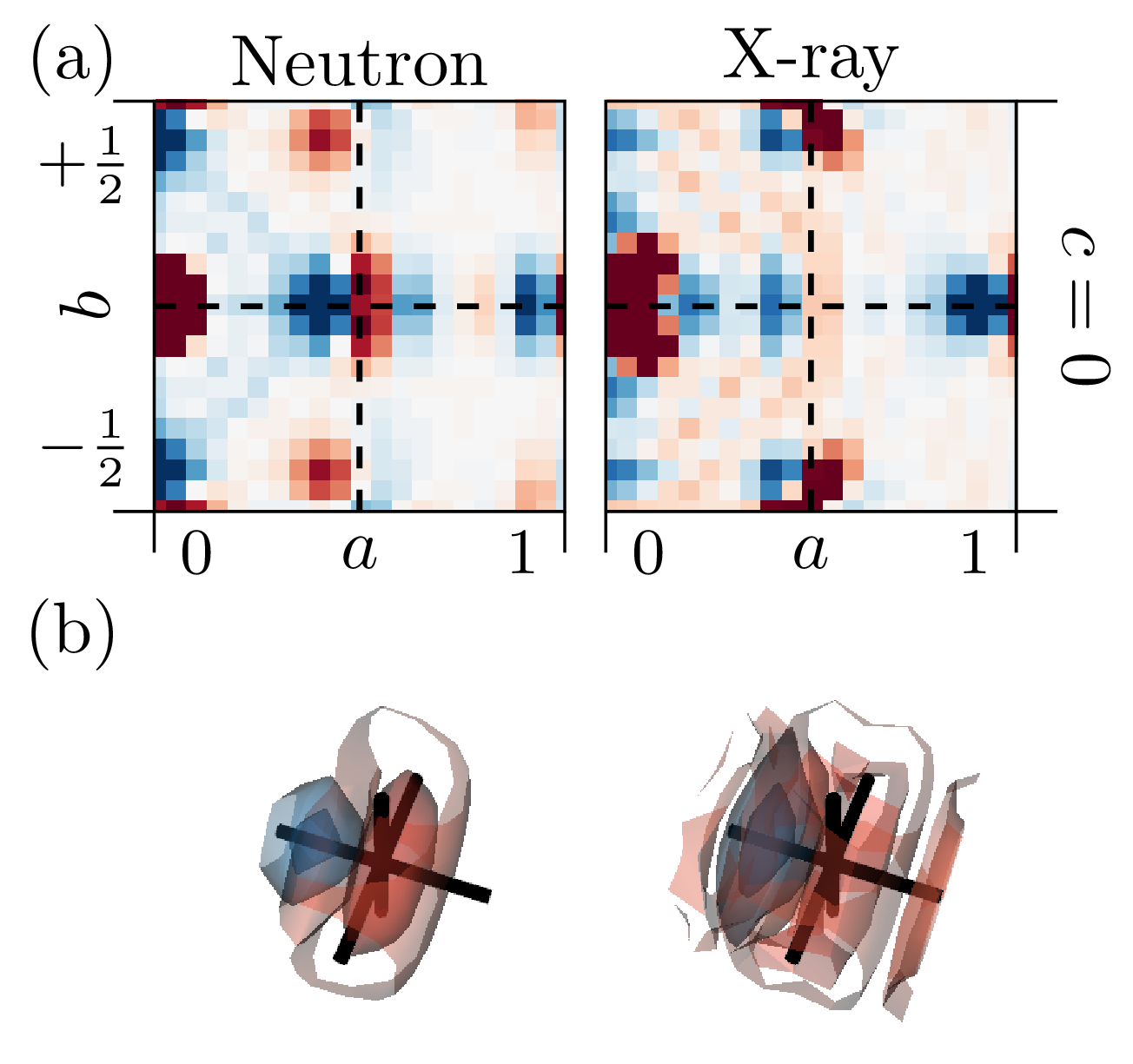}
    \label{Fig:OOdist}
\end{figure}

The  $\langle \frac{1}{2},0,0 \rangle$ interatomic vector is the shortest interatomic vector in the idealized zirconia structure that only occurs in the oxygen sublattice. 
Detailed two-dimensional sections and three-dimensional renderings of the 3D-$\Delta$PDFs around the $(\frac{1}{2},0,0)$ interatomic vector are shown in \Cref{Fig:OOdist}.

For the neutron diffraction experiments, a clear minimum shifted to lower interatomic distances by $\delta^{-}_{\mathrm{OO}}$ at $\left(\frac{1}{2} - \delta^{-}_{\mathrm{OO}}, 0, 0\right)$ and a clear maximum shifted to larger interatomic distances by $\delta^{+}_{\mathrm{OO}}$ at  $\left(\frac{1}{2} + \delta^{+}_{\mathrm{OO}}, 0, 0\right)$ can be observed.  This is the typical signature for a size-effect like relaxation \cite{weber2012three} and a clear indication that locally Oxygen ions relax towards neighbouring vacancies along the $\langle 1 , 0, 0\rangle$ direction.
In the X-ray diffraction experiment the observed trend is much weaker due to the much lower scattering power of oxygen as compared to the heavier metal ions.

This finding is consistent with computational and experimental reports from literature - qualitatively displayed in \Cref{Fig:Relaxation} \cite{frey2005diffuse, asakib1998cation, fevre2005local}. 
The MD simulations of \citeasnoun{fabris2002stabilization} (Zr$_{0.9375}$Y$_{0.0675}$O$_{1.96875}$) report a relaxation of oxygen ions neighbouring a vacancy of 0.27~\AA, the MD simulations of \citeasnoun{fevre2005local}  (Zr$_{0.865}$Y$_{0.135}$O$_{1.9325}$) report a shift of 0.40~\AA \ and the first principle calculations of \citeasnoun{stapper1999ab}  (Zr$_{0.9375}$Y$_{0.0675}$O$_{1.96875}$) report a shift of 0.24~\AA.
Experimental reports using Bragg data refinements from \citeasnoun{goff1999defect} (Zr$_{0.8}$Y$_{0.2}$O$_{1.9}$, shift 0.04 r.l.u. $\approx$ 0.20~\AA),  \citeasnoun{kaiser2005anion} (Zr$_{0.74}$Y$_{0.26}$O$_{1.87}$, shift 0.24~\AA) and  \citeasnoun{ishizawa1999synchrotron} (Zr$_{0.758}$Y$_{0.242}$O$_{1.879}$, shift 0.31~\AA) report similar results.
These results show no clear indication of a correlation of the shift magnitude and the dopant concentration of the sample.
Our results also support a clear shift of the oxygen ions along $\langle 1, 0, 0 \rangle$, and show no indication for previously suggested shifts along $\langle1, 1, 1\rangle$ directions \cite{ishizawa1999synchrotron, argyriou1996neutron} as such a shift would distort the homogeneous maximum at $\left(\frac{1}{2} + \delta^{+}_{\mathrm{OO}}, 0, 0\right)$.
However, an additional, more isotropic off-axis relaxation is possible, as the observed maximum in the 3D-$\Delta$PDF shows a disc like feature.
This indicates that for two oxygen ions that are separated by $\left(\frac{1}{2} + \delta^{a}_{OO} , \delta^{b}_{OO} ,\delta^{c}_{OO} \right)$, $\delta^{a}_{OO}$ shows a much more narrow distribution than $\delta^{b}_{OO}$ and 
$\delta^{c}_{OO}$.

The 3D-$\Delta$PDF analysis also allows a quantitative estimation of the shift magnitudes.
For this purpose we fit the position of the minimum centred at  $\left(\frac{1}{2} - \delta^{-}_{\mathrm{OO}}, 0, 0\right)$ with a three dimensional Gaussian distribution.
The resulting parameter for  $\delta^{-}_{\mathrm{OO}}$ is $1.02(1)\cdot 10^{-1}$ r.l.u. for the fit to the neutron data and $0.91(1)\cdot 10^{-1}$ r.l.u. for the fit to the X-ray data. 
From electrostatic considerations we assume that neighbouring vacancies are highly unlikely. Therefore we interpret the position of the minimum as the average O-vacancy vector. 
With the experimentally obtained lattice parameter $a=5.1505(5)$~\AA\ we estimate that the oxygen ions neighbouring a vacancy relax 0.515(5)~\AA\ along $\langle 1, 0,0 \rangle$ towards the vacancy. 
The shift amplitude is larger than previous experimental reports that utilized Bragg data analysis suggested and more similar to the MD simulations of \citeasnoun{fevre2005local}. 
Due to the much higher relative sensitivity of the neutron 3D-$\Delta$PDF we used the parameter derived from the neutron data in the shift estimation.
Nevertheless, the X-ray 3D-$\Delta$PDF shows a clear signature at $\langle \frac{1}{2},0,0 \rangle$ and the quantitative analysis suggests that the oxygen displacements can be directly observed in the X-ray diffraction experiments, with a reasonable quantitative agreement to the neutron data.

\subsection{ Oxygen metal interactions}

\begin{figure}
    \caption{3D-$\Delta$PDFs obtained from neutron (left) and X-ray (right) diffraction experiments. (a) Two-dimensional section in the $ab0.25$-layer. (b) Three-dimensional rendering of the intensity distribution. Volume shown in the region  $0.1 \le a,b,c \le 0.40$. Black lines indicate the average interatomic distance at $\left( \frac{1}{4}, \frac{1}{4}, \frac{1}{4} \right)$. Positive intensities in red, negative intensities in blue.}
    \includegraphics[width=9cm]{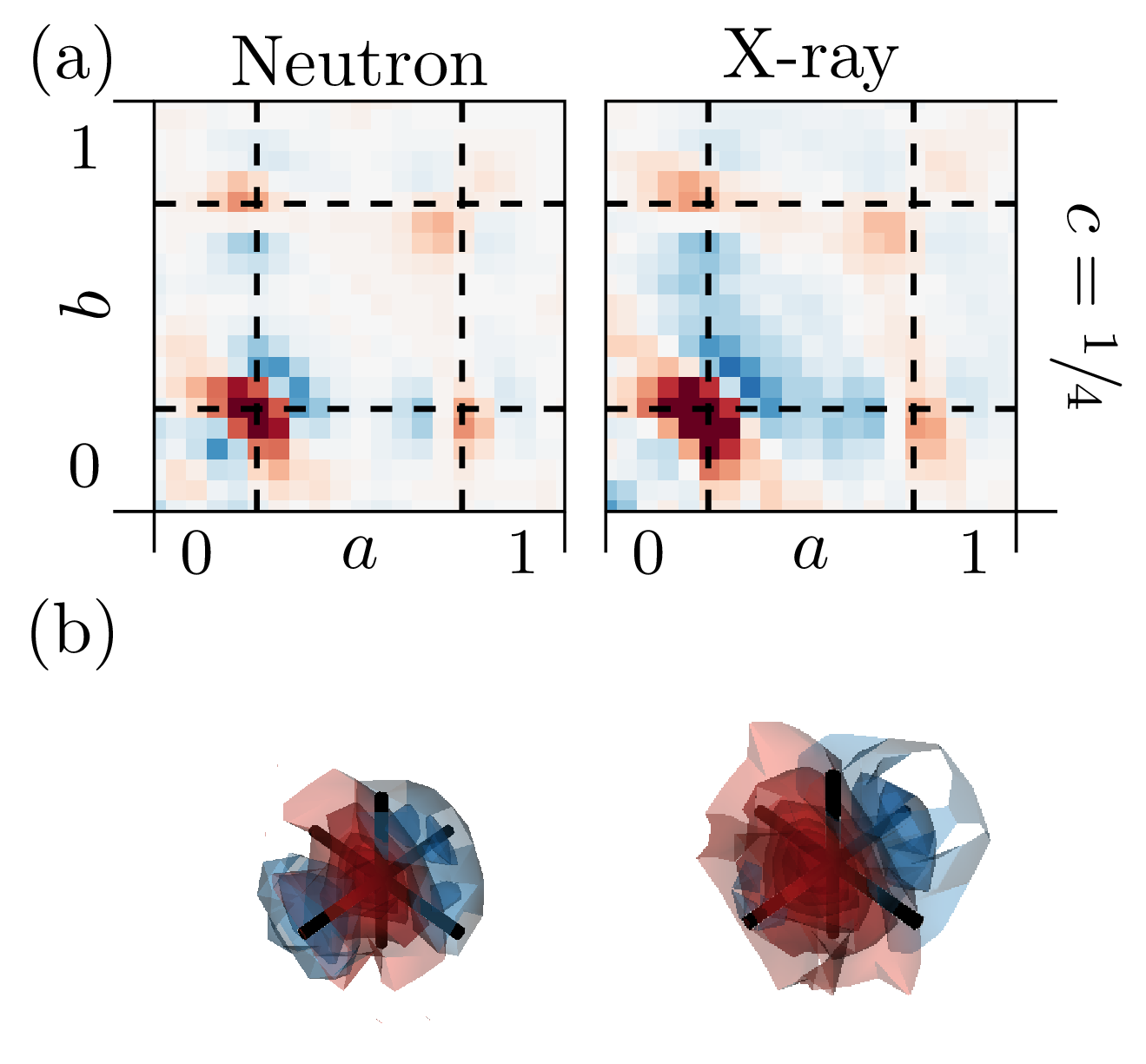}
    \label{Fig:OMdist}
\end{figure}

The  $\langle \frac{1}{4}, \frac{1}{4}, \frac{1}{4} \rangle$ interatomic vectors only occur between the metal and the oxygen sublattice. Detailed two- and three-dimensional 3D-$\Delta$PDFs of the $\left( \frac{1}{4}, \frac{1}{4}, \frac{1}{4} \right)$ interatomic vector are shown in \Cref{Fig:OMdist} and resemble the typical signature for a positive atomic displacement parameter (ADP) correlation: the local bond distance variation is smaller than suggested by the ADPs of the average structure.
The simple analysis and interpretation of interatomic distance as for the $\left(\frac{1}{2}, 0, 0\right)$-vector is not possible here, as there are four distinct pair-correlations that can be observed at the $\langle \frac{1}{4} \ \frac{1}{4} \  \frac{1}{4} \rangle$ interatomic vector:
Possible pairs are: Zr-O, Zr-vacancy, Y-O and Y-vacancy.

For X-ray diffraction, there is negligible contrast between Zr$^{4+}$ and Y$^{3+}$. For neutron diffraction the contrast between Zr and Y is not large (b(Zr) = 7.16 vs. b(Y) = 7.75) either, which does not allow for a direct disentanglement of signatures associated with Y-O and Zr-O distances. Therefore, we interpret the position of the maximum in the 3D-$\Delta$PDF as the average metal-oxygen interatomic vector. 
In a similar fashion to the $\langle \frac{1}{2}, 0, 0 \rangle$ interatomic distance we fit the position of the maximum with a three dimensional Gaussian distribution centred at $\left( \frac{1}{4}+\delta_{\mathrm{MO}}, \frac{1}{4}+\delta_{\mathrm{MO}}, \frac{1}{4}+\delta_{\mathrm{MO}}\right)$. 
The resulting parameters for  $\delta_{\mathrm{MO}}$ are $-1.58(10)\cdot 10^{-2}$ r.l.u. for X-ray and $-0.95(8)\cdot 10^{-2}$ r.l.u. for neutron diffraction.
The observed maximum in the 3D-$\Delta$PDF is shifted towards the centre of the 3D-$\Delta$PDF space, contracted in the direction of the shift and elongated perpendicular of to the shift.
This indicates a local contraction of the oxygen-metal bond length, which can be achieved if metal ions relax away from neighbouring vacancies along the $\langle 1,1,1\rangle$ direction.
The resulting local configurations show a variety of oxygen-metal bond distances, which accounts for the elongation of the maximum perpendicular to the shift direction.
With the experimentally obtained lattice parameter $a=5.1505(5)$~\AA \ we estimate the average metal-O$^{2-}$ nearest neighbour distance at 2.09(1)~\AA\ for X-ray and 2.15(1)~\AA\ for neutron diffraction experiments.
The differences in the calculated distances can be attributed to the differences in scattering contrasts.
The fact that Y has the larger neutron scattering length and the estimated average metal-oxygen bond length is longer for the neutron refinement than for the X-ray refinement indicate that locally Y-O bond lengths are larger than Zr-O bond length, which is consistent with ionic radii~\cite{shannon1976revised,prince2004international}.

The arrangement of the oxygen ions and the vacancies on the oxygen sublattice determines the coordination numbers of the metals.
In the average structure of cubic ZrO$_2$, each Zr$^{4+}$ ion is in regular cubic coordination, while monoclinic  ZrO$_{2}$ only shows 7-fold coordination \cite{frey2005diffuse}.
Y$_{2}$O$_{3}$ shows 6-fold coordinated metal ions at ambient conditions \cite{antic1993structure}.
Hence, it seems that  both the Zr and the Y ions compete for lower coordination numbers; electrostatic effects and atomic sizes need to be taken into account when evaluating which type of metal ions neighbours a vacancy in the real structure.
The simulations of \citeasnoun{bogicevic2003nature} and \citeasnoun{asakib1998cation} suggest that for YSZ NNN vacancies are preferred.
Experimentally this is confirmed by the $^{89}$Y solid-state MAS-NMR studies of \citeasnoun{viefhaus2006solid}, which only show a significant onset of the peaks for 6- and 7-fold coordinated Y-ions above 10~mol\% Y$_{2}$O$_{3}$, which is above the 9~mol\% Y$_{2}$O$_{3}$ in our specimen.

Different metal coordination and different ionic radii lead to different oxygen-metal bond lengths.
In \Cref{Tab:Tab1} we summarize simulated bond-distances by \citeasnoun{asakib1998cation}, experimentally measured bond distances from EXAFS studies in YSZ by \citeasnoun{ishizawa1999synchrotron} and \citeasnoun{catlow1986exafs} and the sum of effective ionic radii based on \citeasnoun{shannon1976revised}.
However, the presence of the oxygen vacancies and the associated relaxations are likely to not yield perfect coordination polyhedra but rather distorted bonding environments (see Figure \ref{Fig:Relaxation}). 
Such distorted bond environments are also supported by density functional theory calculations on superstructures of Zr$_{1-x}$Y$_{x}$O$_{2-x/2}$ \cite{pruneda2005first}, X-ray absorption studies \cite{li1993x} and the fact that 7-fold coordination in pure ZrO$_{2}$ at room temperature shows a considerable variance in  Zr$^{4+}$-O$^{2-}$ bond distances (between 2.05~\AA\ and  2.26~\AA) \cite{yashima1995structural}.

\begin{table}
	\caption{Metal-oxygen bond lengths in \AA\ as a function of coordination number (CN) from MD simulations by \citeasnoun{asakib1998cation}, $^{1}$EXAFS studies by \citeasnoun{ishizawa1999synchrotron} and $^{2}$\citeasnoun{catlow1986exafs} and sum of effective ionic radii based on \citeasnoun{shannon1976revised}.  The simulation of \citeasnoun{asakib1998cation} distinguishes between oxygen vacancy NN to dopant ion and oxygen vacancy NNN to dopant ion. We interpret this in terms of coordination numbers.  $^{a}$NN dopant-vacancy,  $^{b}$NNN dopant-vacancy.   }
	\begin{tabular}{llccc}
		Metal bond & CN & MD & EXAFS & ionic radii\\
		
		\hline
		Y - O &  6 &  2.267$^{a}$  & 2.32$^{1}$ & 2.28 \\
		&  7&  2.267$^{a}$  &  - &  2.34 \\
		&  8&  2.340$^{b}$  & 2.28$^{2}$  & 2.40 \\
		\hline
		Zr - O &  6 &  2.117$^{b}$   &   & 2.10 \\
		&  7&  2.117$^{b}$ & 2.13$^{3}$/2.11$^{2}$  & 2.16 \\
		&  8& 2.119$^{a}$  & 2.11$^{2}$  & 2.22 \\
	\end{tabular}
	\label{Tab:Tab1}
\end{table}

The estimated average metal-oxygen nearest neighbour distances from our 3D-$\Delta$PDF analysis therefore is likely also an average obtained from several distorted bond environments. Comparing the deduced average distance with the sum of the atomic radii \cite{shannon1976revised} in Table~\ref{Tab:Tab1} a 7-fold Zr coordination is supported.
This is consistent with the EXAFS studies by  \citeasnoun{ishizawa1999synchrotron} and \citeasnoun{catlow1986exafs}, the $^{89}$Y solid-state MAS-NMR studies of \citeasnoun{viefhaus2006solid}, as well as the computational work of \citeasnoun{bogicevic2003nature} and \citeasnoun{asakib1998cation} that favour dopant ions as NNN to the vacancies, which results in a preference for Zr in 7-fold coordination.

\subsection{ Metal metal interactions}
\label{sec:MM}
    
      \begin{figure}
    	\caption{3D-$\Delta$PDFs obtained from neutron (left) and X-ray (right) diffraction experiments. (a) Two-dimensional plot in the $ab0$-layer. (b) Three-dimensional rendering of the intensity distribution. Volume shown in the region  $0.25 \le a,b,c+0.5 \le 0.75$. Black lines indicate the average interatomic distance at $\left( \frac{1}{2}, \frac{1}{2}, 0\right)$. Positive intensities in red, negative intensities in blue.}
    	\includegraphics[width=9cm]{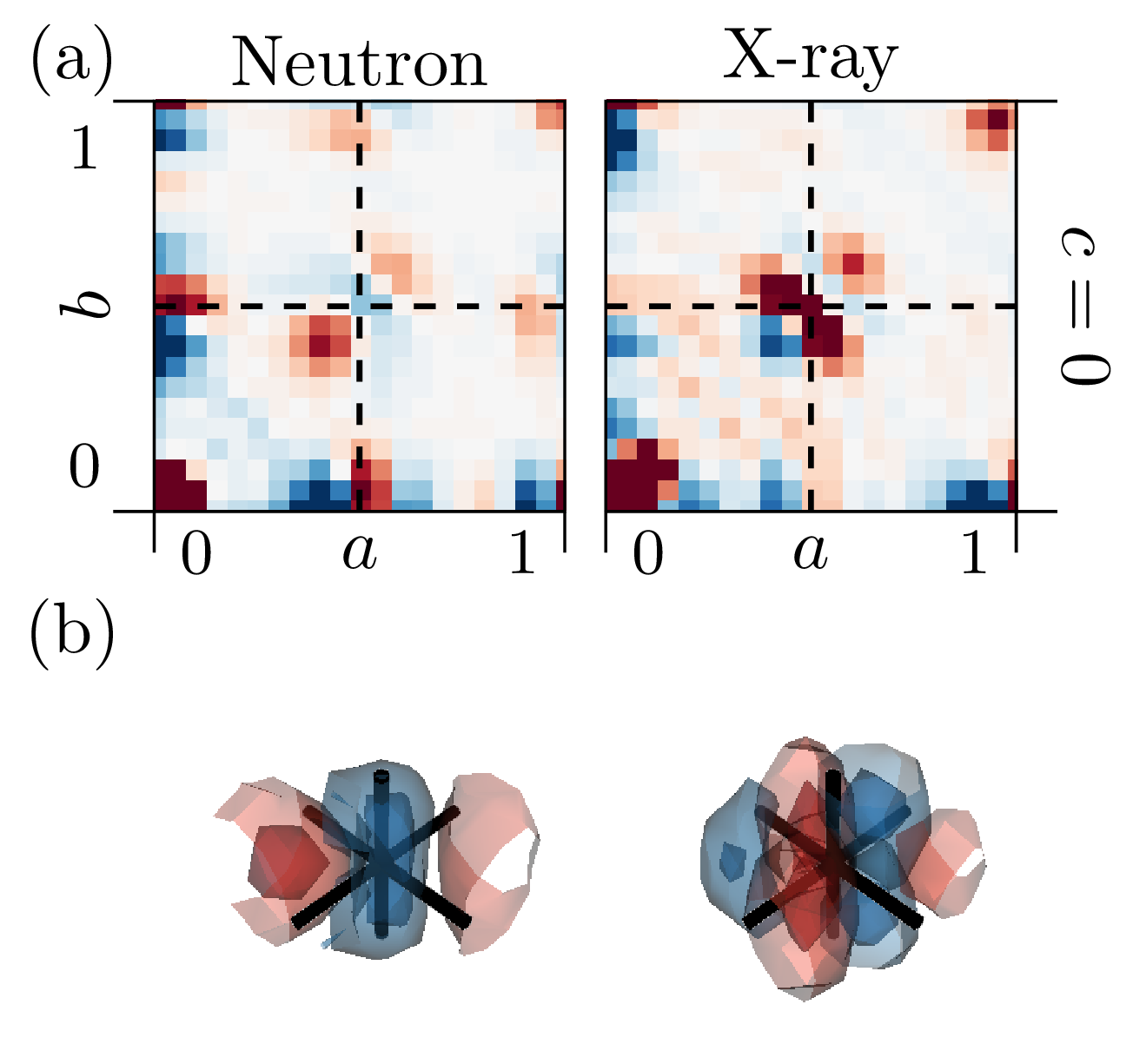}
    	\label{Fig:MMdist}
    \end{figure}
    
    The  $\langle \frac{1}{2}, \frac{1}{2}, 0 \rangle$ is the most complex of the nearest neighbour interatomic distance as it occurs in both sublattices.  The maxima associated with O-O correlations are overlaid by the metal-metal interactions. 3D-$\Delta$PDFs of the $\left( \frac{1}{2}, \frac{1}{2},  0 \right)$ interatomic vector are shown in \Cref{Fig:MMdist}.
    
   For both our diffraction experiments, the observed signatures are complex and the differences highlight the contrast between X-ray and neutron diffraction experiments. However, both signatures show two distinct locations of maxima, which can interpreted as a result of different local configurations:
   configurations where both or one of the bridging oxygen ions are missing in-between the metals will result in a different metal-metal interatomic vector than those of configurations where both bridging oxygen ions are present. 
   Furthermore the exact intensity and distance distribution will depend on the coordination numbers and types of metals involved.
   The entanglement of the interatomic distances involved in the signature at the $\langle \frac{1}{2}, \frac{1}{2}, 0 \rangle$ vector is therefore effectively a multi body correlation and can only be accessed indirectly with scattering experiments where only pair-correlations can be probed directly \cite{welberry2004diffuse, baake2009kinematic}. 
   This limitation is also present in the 3D-$\Delta$PDF and a direct interpretation of the signature at $\langle \frac{1}{2}, \frac{1}{2}, 0 \rangle$ relies on information deduced from shorter interatomic vectors.
 
  However, assuming a shift of the metal ions neighbouring a vacancy away from the vacancy - as deducted from our data in the previous section, we can assign the positive maximum observed at $\left(\frac{1}{2} +\delta_{\mathrm{MM}}, \frac{1}{2} + \delta_{\mathrm{MM}}, 0\right)$: 
  for simplicity we consider a single pair of metal ions M1 at $\left(0,0,0\right)$ and M2 at $\left(\frac{1}{2},\frac{1}{2},0\right)$ with the bridging vacancy located at $\left(\frac{1}{4},\frac{1}{4},\frac{1}{4}\right)$. The metal ions will relax away form the vacancies in the $\langle 1,1,1\rangle$ direction. M1 will be shifted to $\left(-\Delta_{\mathrm{MO}},-\Delta_{\mathrm{MO}},-\Delta_{\mathrm{MO}}\right)$ and M2 will be shifted to $\left(\frac{1}{2} + \Delta_{\mathrm{MO}},\frac{1}{2}+\Delta_{\mathrm{MO}},-\Delta_{\mathrm{MO}}\right)$, yielding the new inter-atomic vector between M1 and M2 as $\left(\frac{1}{2} + 2\Delta_{\mathrm{MO}},\frac{1}{2} + 2\Delta_{\mathrm{MO}} ,0\right)$.
 While the shift $\delta_{\mathrm{MO}}$ determined in the previous section yields an average metal-oxygen bond length, the shift $\Delta_{\mathrm{MO}}$ we determine here is directly related to the shift of a metal atom neighbouring a vacancy.
 We fit the shift in the X-ray diffraction 3D-$\Delta$PDF as the scattering contributions here are dominated by the contributions of the metal ions.
The fitted resulting shifts is $\delta_{\mathrm{MM}} = 1.043(4)\cdot10^{-1}$ r.l.u.
This corresponds to a metal shift along the $\langle 1, 1, 1\rangle$ direction away from the vacancy of 0.269(2)~\AA, which is slightly more pronounced than references in the literature.
The MD simulations of \citeasnoun{fabris2002stabilization} (Zr$_{0.9375}$Y$_{0.0675}$O$_{1.96875}$) report a metal shift away from the a neighbouring a vacancy of 0.18~\AA, the MD simulations of \citeasnoun{fevre2005local}  (Zr$_{0.865}$Y$_{0.135}$O$_{1.9325}$) report a shift of 0.1~\AA \ and the first principle calculations of \citeasnoun{stapper1999ab}  (Zr$_{0.9375}$Y$_{0.0675}$O$_{1.96875}$) report a shift of 0.18~\AA.
Experimental reports using Bragg data refinements from \citeasnoun{goff1999defect} (Zr$_{0.8}$Y$_{0.2}$O$_{1.9}$, shift 0.028 r.l.u. $\approx$ 0.11~\AA),  \citeasnoun{kaiser2005anion} (Zr$_{0.74}$Y$_{0.26}$O$_{1.87}$, shift 0.22~\AA) and  \citeasnoun{ishizawa1999synchrotron} (Zr$_{0.758}$Y$_{0.242}$O$_{1.879}$, 0.0219 r.l.u. shift 0.08~\AA) vary largely in the reported shift magnitude and show no clear indication of a correlation of the shift magnitude and the dopant concentration of the sample.

 \subsection{Possibility of vacancy clustering and interstitial metal atoms at  $\left( \frac{1}{2}, \frac{1}{2}, \frac{1}{2} \right)$ }
 
\begin{figure}
 	\caption{3D-$\Delta$PDFs obtained from neutron (left) and X-ray (right) diffraction experiments. (a) Two-dimensional plot in the $ab0.5$-layer. (b)Three-dimensional rendering of the intensity distribution. Volume shown in the region  $0.25 \le a,b,c \le 0.75$. Black lines indicate the average interatomic distance at $\left( \frac{1}{2}, \frac{1}{2}, \frac{1}{2}\right)$. Positive intensities in red, negative intensities in blue.}
 	\includegraphics[width=9cm]{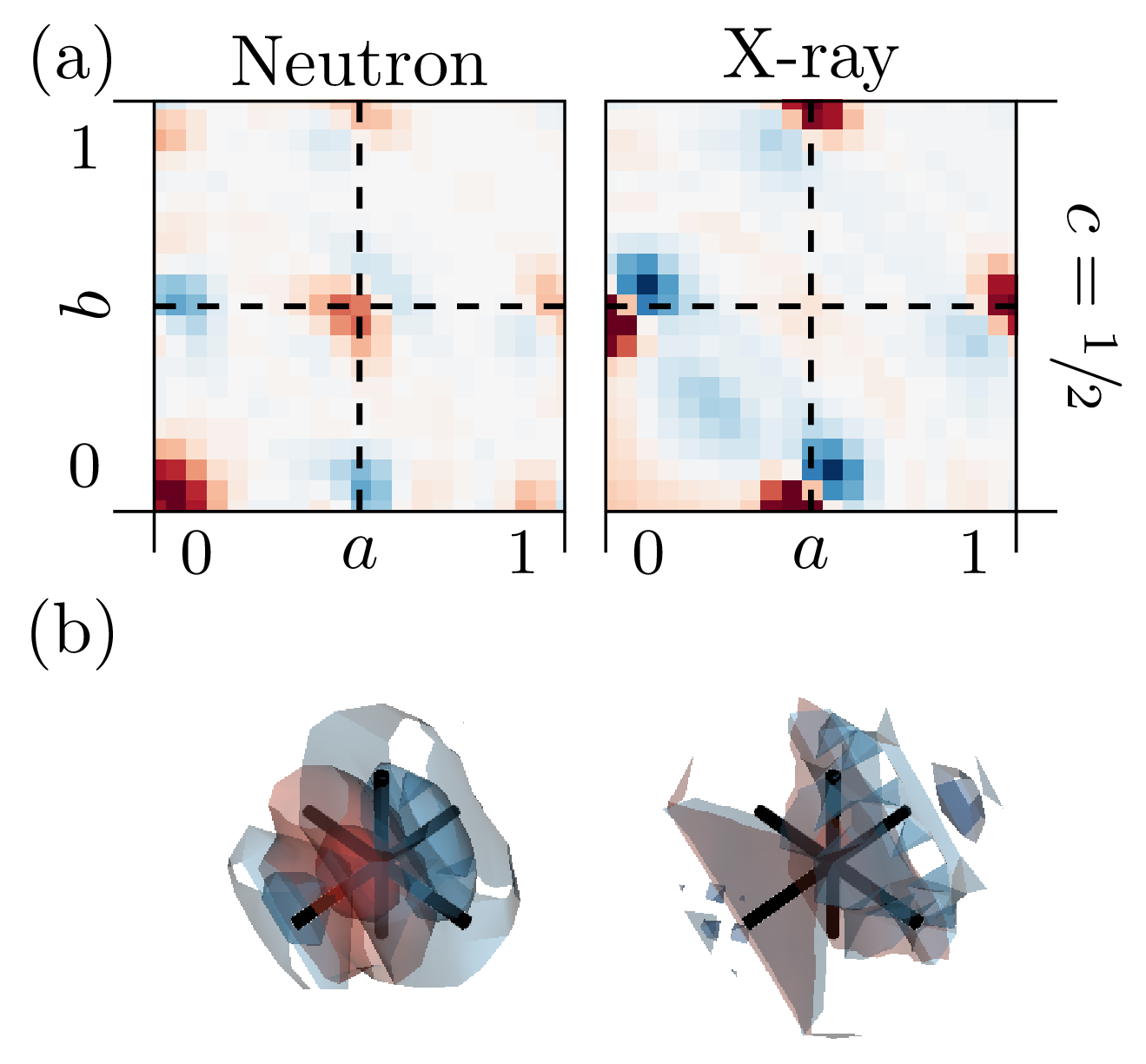}
 	\label{Fig:DoubleVac}
 \end{figure}

 Questions that have been discussed controversially in the literature are the possibility of vacancy clusters along the  $\langle \frac{1}{2}, \frac{1}{2}, \frac{1}{2} \rangle$ vector - eventually forming chains of pyrochlore like structural elements \cite{welberry19933d,goff1999defect} - and the possibility of interstitial metal ions at $\left( \frac{1}{2}, \frac{1}{2}, \frac{1}{2}\right)$ \cite{goff1999defect}.
 Both of these questions can be clarified by analysing the 3D-$\Delta$PDFs in the vicinity of $\langle \frac{1}{2}, \frac{1}{2}, \frac{1}{2} \rangle$ - see Figure \ref{Fig:DoubleVac}.
 
 The possibility of metal ion interstitials at  $\left( \frac{1}{2}, \frac{1}{2}, \frac{1}{2}\right)$ as considered by  \cite{goff1999defect} can also be ruled out here by the absence of relevant signatures in the X-ray 3D-$\Delta$PDF in \Cref{Fig:DoubleVac}. 
 Possible interstitial metal ions at $\left( \frac{1}{2}, \frac{1}{2}, \frac{1}{2}\right)$ would lead to metal-metal interatomic vectors at $\left( \frac{1}{2}, \frac{1}{2}, \frac{1}{2}\right)$, which are not described by our average structural model and would hence result in a signature in the 3D-$\Delta$PDF.
 The lower scattering power of oxygen compared to the metal ions - supported by the relative strength of the signatures in the X-ray 3D-$\Delta$PDFs observed at $\langle \frac{1}{2}, 0, 0\rangle$ and $\langle \frac{1}{2}, \frac{1}{2}, 0 \rangle$ - suggests that the X-ray 3D-$\Delta$PDF would  be very sensitive to such a signature, while in the neutron case the signature present at $\langle \frac{1}{2}, \frac{1}{2}, \frac{1}{2} \rangle$ can be interpreted in terms of oxygen-oxygen correlations.
 
 The shortest oxygen-oxygen interatomic vectors are $\langle \frac{1}{2}, 0, 0 \rangle$ and $\langle \frac{1}{2}, \frac{1}{2}, 0 \rangle$. For these vectors vacancy clustering becomes highly unlikely from an electrostatic viewpoint \cite{bogicevic2003nature}. 
 The shortest vector where vacancies are allowed from electrostatic considerations is the $\langle \frac{1}{2}, \frac{1}{2}, \frac{1}{2} \rangle$ vector. This yields the possibility to obtain 6-coordinated metal ions in the structure. 
  The doping level in the specimen at hand (Zr$_{0.82}$Y$_{0.18}$O$_{1.82}$) is low enough to distribute the vacancies in the structure without needing to form 6-fold coordinated ions, unlike for higher doped variants such as described in \citeasnoun{welberry1992diffuse}. The in-depth simulation based study of \citeasnoun{bogicevic2003nature} however suggests that the formation of double vacancies along $\langle \frac{1}{2}, \frac{1}{2}, \frac{1}{2} \rangle$  can nevertheless be favourable. 
    
  The signatures in \Cref{Fig:DoubleVac} are very weak in the case of X-ray diffraction - basically indistinguishable from residual background noise.
  The situation is different for the neutron scattering case: here the scattering length of oxygen is comparable to that of the metals and the relative strength of signatures attributed ot oxygen-oxygen interactions is much greater than in the case of X-ray diffraction. 
  Here, we observe a clear positive signature at $\left(\frac{1}{2}, \frac{1}{2}, \frac{1}{2}\right)$  suggesting a positive correlation, i.e. a tendency to form double-vacancy pairs along this interatomic distance.
  Comparing the strength of the signature with the signature at $\langle \frac{1}{2}, 0, 0 \rangle$ (note \Cref{Fig:OOdist} and \Cref{Fig:DoubleVac} are on the same relative scale), the signature of the double vacancies along $\langle \frac{1}{2}, \frac{1}{2}, \frac{1}{2} \rangle$  is much weaker.
  This suggests that there is no strict formation of double vacancies which would in turn lead to formation of zigzag-chains of 6-coordinated metal ions and local pyrochlore-like structural elements.
  This observation is in agreement with the findings of \citeasnoun{welberry1995modulation}.

  \subsection{Longer range interactions and extent of correlations}
   
   A direct interpretation of longer range interactions in terms of underlying configurations is basically impossible due to the configurational complexity of possible arrangements of nearest neighbour, next nearest neighbour and next next nearest neighbour oxygen-vacancy pairs and metal-metal pairs.
   Nevertheless, the obtained 3D-$\Delta$PDFs in Figure \ref{Fig:RealSpaceMaps} allow a direct judgement of the extent to which correlations due to the previously identified local correlations persist - for both the neutron and X-ray 3D-$\Delta$PDF we do not observe correlations that are significantly longer than two unit cells, i.e. $\approx 10$~\AA. 
   This is significantly shorter than previously reported in literature \cite{goff1999defect,frey2005diffuse}.
   
  Possible differences in the neutron and X-ray 3D-$\Delta$PDFs allow distinguishing whether the correlations are propagated in both sublattices to the same extent: the limited sensitivity of X-ray 3D-$\Delta$PDFs to oxygen-oxygen correlations allows the interpretation of the extent of correlations in terms of metal displacements only.
  In the X-ray 3D-$\Delta$PDF correlations along $\langle 1, 1, 0\rangle$ persist longer than in the neutron case (see e.g. signature at $\langle \frac{3}{2}, \frac{3}{2}, 0\rangle$ in \Cref{Fig:RealSpaceMaps}) - providing evidence that the metal-metal nearest neighbour distortion described in \Cref{sec:MM} is propagated to further neighbours in the same direction.
 
\subsection{3D-$\Delta$PDF informed modelling}
\label{sec:model}

The detailed analysis of the experimentally obtained 3D-$\Delta$PDFs we performed in the previous sections lays the groundworks for simulating a simplistic atomistic model that realizes the correlations we derived. 
For this purpose we utilize three consecutive Monte Carlo simulations where the first two simulations establish the chemical ordering and the third simulation relaxes the atomistic positions in accordance with our analysis.

For the simulations we utilize five model crystals of $10\times10\times10$ unit cells. First the metal ions are distributed at random to match the average composition of 82~\% Zr and 18~\% Y ions. Charge balance is ensured by removing the suitable amount of oxygen ions.

The first Mote Carlo simulations induces chemical ordering of the metal ions. We assume for electrostatic reasons that Y ions tend to avoid being close to each other. The Monte Carlo simulations swaps Zr and Y ions to avoid NN ad NNN Y-Y pairs.

The second Monte Carlo simulation then swaps oxygen ions and vacancies in such a fashion that:
\begin{enumerate}
	\item Y-ions prefer 8 fold coordination and avoid being next to vacancies,
	\item Vacancy pairs separated by $\langle \frac{1}{2}, 0, 0 \rangle$ and $\langle \frac{1}{2}, \frac{1}{2}, 0 \rangle$ are penalized with a high energy,
	\item a minimal energy gain is introduced for vacancies separated by $\langle \frac{1}{2}, \frac{1}{2}, \frac{1}{2} \rangle$.
\end{enumerate}
The resulting structures fulfil (1) and (2) without violations. On average 3.06(13)~\% of the metal ions  are 6-fold coordinated.

The third Monte Carlo simulation then introduces lattice relaxations as deduced form the analysis of the 3D-$\Delta$PDFs:
Assuming that every vacancy introduced in the structure causes its six neighbouring oxygen ions to relax towards the vacancy along the $\langle 1, 0, 0 \rangle$ direction and its four nearest neighbour metal ions to relax along $\langle 1, 1, 1 \rangle$ direction away from the vacancy, we introduced a static shift for the respective percentage of the ions and then used the Monte Carlo algorithm as implemented in the DISCUS program \cite{neder2008diffuse} to switch the displacements.
The energy targets were set as spring potentials with metal-oxygen target distances taken as the sum of the ionic radii \cite{shannon1976revised}.

\begin{figure}
	\caption{3D-$\Delta$PDFs obtained from our simplified atomistic model (M) from neutron (left) and X-ray (right) diffuse scattering calculations compared to the experimentally obtained 3D-$\Delta$PDFs (E). (a)  Two-dimensional section around $\left(\frac{1}{2}, 0, 0 \right)$ in the $ab0$-layer, model top, experiment bottom.  (b)  Two-dimensional section around $\left(\frac{1}{4}, \frac{1}{4}, \frac{1}{4} \right)$ in the $ab0.25$-layer,  model top left, experiment bottom right. (c)  Two-dimensional section around $\left(\frac{1}{2}, \frac{1}{2}, 0\right)$ in the $ab0$-layer model top left, experiment bottom right. (b)  Two-dimensional section around $\left(\frac{1}{2}, \frac{1}{2}, \frac{1}{2} \right)$ in the $ab0.5$-layer, model top left, experiment bottom right. }
	\includegraphics[width=9cm]{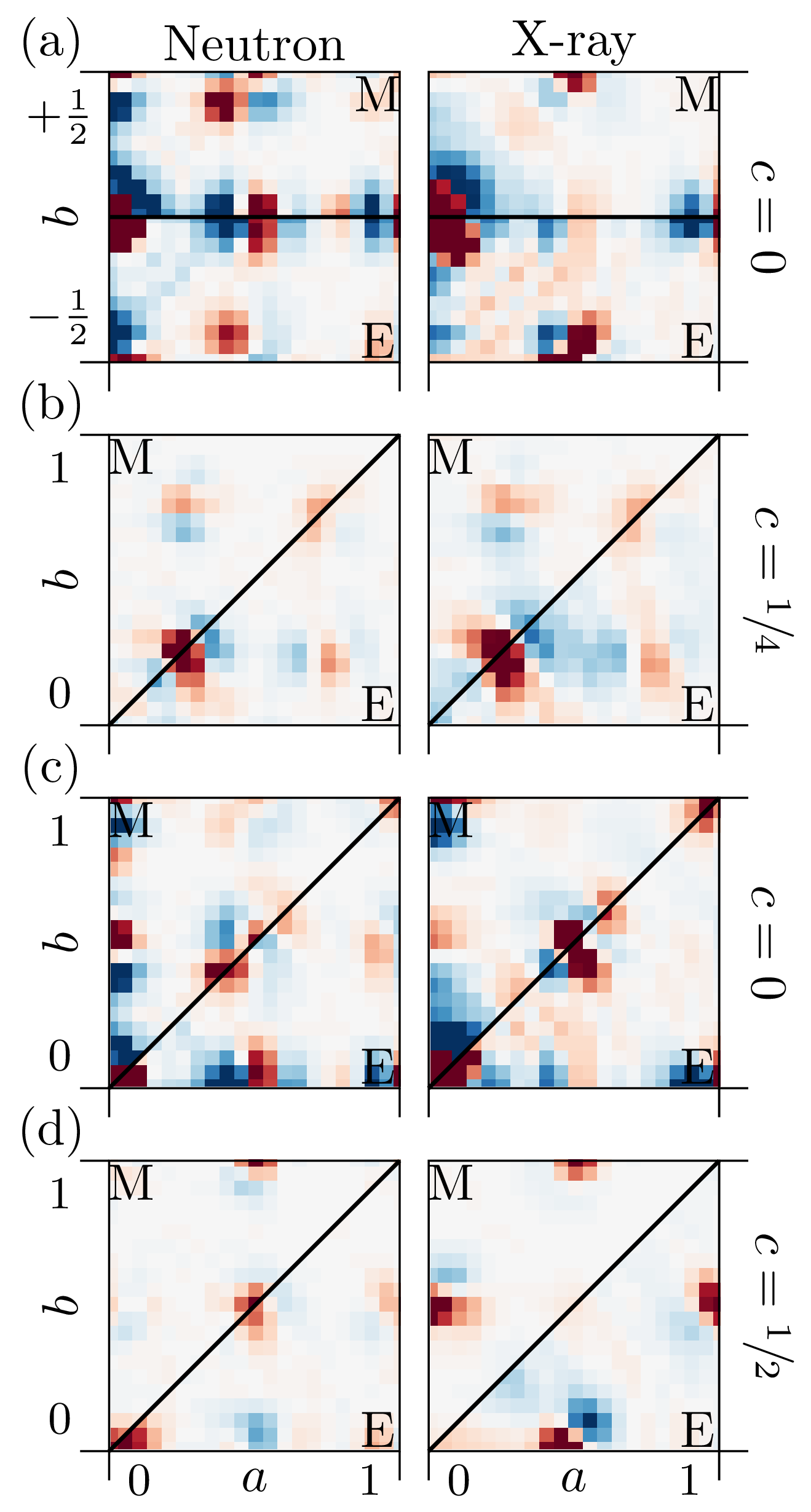}
	\label{Fig:Model}
\end{figure}

The diffuse scattering (see Supporting Information) was calculated using the DISCUS program \cite{neder2008diffuse} on a grid adapted to the size of the supercell. To obtain the model 3D-$\Delta$PDFs we utilized a customized punch and fill algorithm (punch size was one voxel, as there is no experimental broadening of Bragg reflections in the calculated diffuse scattering).
The same Gaussian falloff was multiplied to the simulated data as was to the experimental data and Meerkat \cite{meerkat} was used to obtain the resulting simulated 3D-$\Delta$PDFs in the regions of interest shown in Figure \ref{Fig:Model} and in the supporting information.

A comparison of the 3D-$\Delta$PDFs from our simple model and the respective sections in the experimentally obtained 3D-$\Delta$PDFs in Figure \ref{Fig:Model} shows that the nature of the major features that we quantified in our analysis are reproduced. The majority of the positions of maxima and minima are reproduced well while we observe more significant differences in the intensity distribution and the spatial extent of the features.
These discrepancies we attribute to the simplicity of our model that only contains the quantities derived from a direct interpretation of the experiment.
A refinement, e.g. using  a reverse Monte Carlo simulation, would likely result in a much better agreement between data and model.
However our emphasis here is on the direct interpretation of the 3D-$\Delta$PDF and our model confirms that the quantities we derive from this direct interpretation can be used to build a simplistic model that reproduces the nature of the local interactions and therefore confirms the results from our direct interpretation.

\section{Concluding Remarks}

The defect structure of YSZ has been previously studied by several computational and experimental techniques \cite{frey2005diffuse}.
Bragg data analysis, NMR and EXAFS studies only yield limited information on the local order in the system and a comprehensive model of the defect structure can only be obtained by the analysis of single crystal diffuse scattering.
In the past such analysis involved computationally expensive modelling and a direct interpretation of present distortions from the measured reciprocal space sections required very involved expert knowledge or was simply impossible~\cite{andersen1986defect,welberry19933d,welberry1995modulation,goff1999defect}.
With the advances in experimental techniques it is now possible to obtain full three-dimensional reciprocal space coverage and the application of the 3D-$\Delta$ PDF enables a direct interpretation of the data in terms of defect models \cite{weber2012three,roth2019solving,simonov2014experimental}.

In our contribution we demonstrated how we use the combined information from X-ray and neutron 3D-$\Delta$PDF to directly establish a quantitative defect model for YSZ. 
The signatures we observe on the shortest interatomic vectors - i.e. $\langle \frac{1}{2}, 0, 0 \rangle$, $\langle \frac{1}{4}, \frac{1}{4}, \frac{1}{4} \rangle$, $\langle \frac{1}{2}, \frac{1}{2}, 0 \rangle$ and $\langle \frac{1}{2}, \frac{1}{2}, \frac{1}{2} \rangle$ - enabled us to build a conclusive model for local correlations.
For the quantitative analysis the combination of the two different radiation types is crucial: the X-ray 3D-$\Delta$PDF has a limited sensitivity to oxygen-oxygen correlations, while in the neutron 3D-$\Delta$PDF the signatures of metal-metal correlations and oxygen-oxygen correlations overlap. The combined analysis enables a disentanglement of the signatures and  permits the quantitative analysis of local relaxation mechanisms. The shifts directions we proposed from our analysis are in good agreement with previously reported relaxations from different experimental investigations and simulation efforts, the shift magnitudes we analyse are more pronounced than previously suggested.. 
We consider our work and important first step in the progress towards to a direct methods type analysis for diffuse scattering using 3D-$\Delta$PDF methods.

Our study shows how a simplistic, direct interpretation of the 3D-$\Delta$PDF leads to results that align well with established interpretations in literature, that involved complex modelling of the described correlations.
In agreement with \cite{goff1999defect} we do not find evidence for interstitial metal atoms at $\left(\frac{1}{2}, \frac{1}{2}, \frac{1}{2}\right)$. 
We find strong evidence for double vacancy pairs along $\langle1/2, 1/2, 1/2\rangle$ \cite{bogicevic2003nature,welberry19933d}.
Our analysis confirms oxygen displacements along $\langle1,0,0\rangle$ and metal displacements along $\langle1,1,1\rangle$ \cite{kaiser2005anion}, while we do not find direct evidence for oxygen displacements along $\langle1,1,1\rangle$ \cite{ishizawa1999synchrotron, argyriou1996neutron} or metal displacements along $\langle1,1,0\rangle$ \cite{welberry19933d}.

We present a simplistic static local order model that describes the most local interatomic correlations well, but not all signatures that can be observed in the data are reproduced - especially the observed longer range correlations are not described in the model presented here.
A possibility to include such longer range interactions would be a big box modelling approach, e.g. a reverse Monte Carlo Simulation: by fixing the directions and the magnitudes of the shifts of the atoms to the values identified in our 3D-$\Delta$PDF analysis a reverse Monte Carlo Simulation would be able to generate a model crystal that captures longer range correlations that are not directly interpretable from the 3D-$\Delta$PDF analysis. In turn this would then be a tool to identify longer range interactions that are not present in the current model.

The measurements and model we obtained here were at ambient conditions.
However, for the technical application of YSZ as an oxygen ion conductor the local order at elevated temperatures may be more relevant \cite{devanathan2006computer,kaiser2005anion,krishnamurthy2004oxygen,itoh2015effect}. 
For variable temperature studies the 3D-$\Delta$PDF analysis as we present here facilitates a direct and quantitative comparison of correlations in real space as a function of temperature and therefore enables a more complete picture than previously analysed selected sections of reciprocal space.

Another aspect that is of importance for the technological application is the interplay of compositional disorder and lattice dynamics. 
The picture we present here only describes a static local order model and hence does not cover the full picture of correlations at ambient conditions.
The possible coupling of the local order to the lattice dynamics of a system is a particular challenge \cite{ziman1979models,snyder2011complex}. In YSZ the phonon anharmonicity at lower temperatures was attributed to the defect structure and in turn to a higher vibrational entropy \cite{li2015phonon}.
A comprehensive, quantitative and reliable local order model is needed for a complete understanding of the lattice dynamics in complex systems such as YSZ.
In the future, the combination of temperature dependent energy discriminated diffraction experiments  and supercell lattice dynamic calculations \cite{overy2017phonon,schmidt2022interplay} based on our local order model will enable a complete understanding of the structure and dynamics of such a complex locally ordered system.

		\ack{Acknowledgements}

	We gratefully acknowledge the financial support of the ERC grant COMPLEXORDER (number 788144), the beam time at ILL D19 (ILL data DOI: 10.5291/ILL-DATA.5-13-277) and Colin Johnston from Oxford University for running and interpreting EDX measurements.
	Nebil Ayape Katcho (ILL) is thanked for the detailed help with the neutron data reconstruction. Arkadiy Simonov (ETH Z\"urich) is acknowledged for providing the Meerkat software. Yahua Liu, Feng Ye, Christina Hoffmann (all Oak Ridge National Laboratory) and Paul Benjamin Klar (University of Bremen) are acknowledged for fruitful discussions.

\referencelist[References]

\end{document}